\documentclass[aps,prb,twocolumn,showpacs,superscriptaddress,groupedaddress]{revtex4-1}
\usepackage{amsmath}
\usepackage{textcomp}
\usepackage{bm}
\usepackage{xcolor}
\usepackage{subfigure}
\usepackage{graphicx}
\usepackage{graphics}
\begin{document}
	
\title{Domain size and charge defects on the polarization switching of antiferroelectric domains}

\author{Jinghao Zhu$^{1}$}
\author{Zhen Liu$^{2,3}$}
\email{liuz_hit@njust.edu.cn}
\author{Boyi Zhong$^{3}$}
\author{Yaojin Wang$^{2}$}
\email{yjwang@njust.edu.cn}
\author{Bai-Xiang Xu$^{3}$}
\email{xu@mfm.tu-darmstadt.de} 
\affiliation{$^1$Nanjing Research Institute Of Electronics Technology, Nanjing, 210039, Jiangsu, China}
\affiliation{$^2$School of Materials Science and Engineering, Nanjing University of Science and Technology, Nanjing 210094, Jiangsu, China}
\affiliation{$^3$Mechanics of Functional Materials, Department of Materials Science, Technical University of Darmstadt, 64287 Darmstadt, Germany}

\date{\today}

\begin{abstract}
The switching behavior of antiferroelectric domain structures under the applied electric field is not fully understood. In this work, by using the phase field simulation, we have studied the polarization switching property of antiferroelectric domains. Our results indicate that the ferroelectric domains nucleate preferably at the boundaries of the antiferroelectric domains, and antiferroelectrics with larger initial domain sizes possess a higher coercive electric field as demonstrated by hysteresis loops. Moreover, we introduced charge defects into the sample and numerically investigated their influence.  It is also shown that charge defects can induce local ferroelectric domains, which could suppress the saturation polarization and narrow the enclosed area of the hysteresis loop. Our results give insights into understanding antiferroelectric phase transformation and optimizing the energy storage property in experiments.

\end{abstract}

\maketitle
\section{Introduction}
Antiferroelectric (AFE) structure was first predicted by Kittel in 1951. Almost at the same time, PbZrO$_3$ was confirmed to be a perovskite AFE at room temperature. Over the past decades, many antiferroelectric materials are successively found in experiments, such as PbHfO$_3$, AgNbO$_3$, NaNbO$_3$ and so on. AFEs could transform to ferroelectric (FE) state under the external electric field, giving rise to the special double hysteresis loop. Such behavior makes AFEs excellent dielectric energy-storage materials for high-density capacitors. Many works have shown that the energy storage performance of AFE materials is not only determined by the chemical composition but also significantly affected by the domain structures of the materials\cite{xu1995impurity,tan2011antiferroelectric,zhuo2018large,wang2019ultrahigh,qi2019ultrahigh,patel2014enhancing}. For example, it is demonstrated that the field-induced multiphase transition of (Pb$_{0.98}$La$_{0.02}$)(Zr$_{0.55}$Sn$_{0.45}$)$_{0.995}$O$_3$ ceramic can lead to an ultrahigh energy storage density\cite{wang2019ultrahigh}. Utilizing nanoscale domains and ferroelastic domain switching can further enhance the energy storage density of AFE ceramics\cite{qi2019ultrahigh,patel2014enhancing}. Therefore, understanding the structure effect and transformation behavior of AFE domains is of great importance to improve the material properties. Experimentally, X-ray diffraction study has been used to study the lattice structure\cite{jona1957x,shirane1953phase,blue1996situ}, and high resolution scanning transmission electron microscope (STEM) has been used to investigate the modulation of AFE incommensurate phases\cite{he2005electric,asada2004induced,guo2015direct}. The atomically resolved STEM images have provided us the atomic cation displacement and polarization arrangement of AFE domain structure and boundaries\cite{fan2020tem,ma2019atomically,ma2019uncompensated,fu2020unveiling}.  Recent findings indicate that the polarization of AFE incommensurate phases varies in magnitude and is not fully compensated\cite{ma2019uncompensated,fu2020unveiling}. The polarization arranges in a quite different manner from the ferroelectrics at the AFE boundaries\cite{ma2019atomically}.
%Theoretical predictions have figured out that the AFE transformation is closely related to the complex coupling between the soft polar mode and oxygen octahedron tilt mode\cite{fthenakis2017dynamics,hlinka2014multiple}.

However, due to the complex subdomain microstructure and difficulties of experimental measurements, several fundamental puzzles still remain. For instance, in the case that AFE and FE domains coexist, the polarization arrangement of FE and AFE interfaces lack investigation. In particular, the polarization vectors rotate and switch in the AFE domain structures under the applied electric field is unclear. Moreover, it is known that the initial domain size and charge defects can greatly affect the dielectric and piezoelectric properties of FEs. How these aspects influence the switching mechanisms and the hysteresis loops of AFEs is still an open question. To address the questions, in this work, we have carried out phase field simulations to study the domain structure and polarization switching property, based on the developed Landau model of AFEs proposed in our previous work\cite{liu2020insight}. In Sec. II, a two-dimensional phase field model of AFEs based on the Landau theory is introduced in order to study the domain structures. In Sec. III, we first study the polarization switching behavior of the AFE domains and the size effect on the hysteresis loop without considering charge defects, and then predict the influence of charge defects on the domain structures and hysteresis loop. The conclusions are given in Sec. IV.

\section{Modeling}

In the Landau model of AFEs, the free energy is expressed as a polynomial in terms of the polarization and the oxygen octahedron tilt angle\cite{liu2020insight}. As we mainly focus on the polarization distribution and switching of the structures, for simplicity, we assume the oxygen octahedron tilt angle as constant, thus, in the two dimensional phase field model the total free energy of the AFEs in terms of the polarization $\mathbf{P}(P_x,P_y)$  can be written as 
\begin{equation}
F=\int (f_L+f_g+f_{ela}-\mathbf{E}_d\cdot\mathbf{P}-\mathbf{E}_{ext}\cdot\mathbf{P}){\rm d}V
\end{equation}
where $f_L$  denotes the classical Landau free energy density of ferroelectrics, and $f_{ela}$ is the elastic free energy density. The term $f_g$ is contributed by the spatial polarization changes that can determine the stability of FE and AFE phases. $E_d$ is the depolarization field and $E_{ext}$ is the external electric field. $f_L$ is given by
\begin{equation}
\begin{split}
f_L=&\alpha_1(P_x^2+P_y^2)+\alpha_{11}(P_x^4+P_y^4)+\alpha_{111}(P_x^6+P_y^6)\\
&+\alpha_{12}P_x^2P_y^2+\alpha_{112}(P_x^2P_y^4+P_x^4P_y^2)
\end{split}
\end{equation}
where $\alpha_1,\alpha_{11},\alpha_{111}, \alpha_{12}$, and $\alpha_{112}$ are the Landau free energy coefficients. The term $f_g$ could be regarded as gradient energy which contains first order and second-order derivative of the polarization.
\begin{equation}
\begin{split}
	f_g=\sum_i \left\{\lambda\left[(\frac{\partial P_i}{\partial x})^2+(\frac{\partial P_i}{\partial y})^2\right]+g\left[(\frac{\partial^2 P_i}{\partial x^2})^2+(\frac{\partial^2 P_i}{\partial y^2})^2\right]\right\}
\end{split}
\end{equation}
where $i=x,y$. The coefficients $\lambda$ and $g$ ($g>0$) are dependent on the oxygen tilt angle. For $\lambda<0$, the AFE phases are more stable. Otherwise, the FE state is favored. The ratio between $\lambda$ and $g$ could determine the modulation of the AFE phases. The elastic free energy density of the two dimensional system is expressed as
\begin{equation}
\begin{split}
f_{ela}=&\frac{1}{2}C_{11}\left[(\epsilon_{11}-\epsilon_{11}^0)^2+(\epsilon_{22}-\epsilon_{22}^0)^2 \right] \\
&+C_{12}(\epsilon_{11}-\epsilon_{11}^0)(\epsilon_{22}-\epsilon_{22}^0) \\
&+2C_{44}(\epsilon_{12}-\epsilon_{12}^0)^2
\end{split}
\end{equation}
where $\epsilon_{ij}$ represents the total elastic strain, and  $\epsilon_{ij}^0$ is the spontaneous strain which is related to the spontaneous polarization and oxygen tilt. $\epsilon_{ij}^0=Q_{ijkl}P_kP_l+q_{ijkl}\theta_k\theta_l$, in which $\theta_i$ represents the oxygen tilt angel. For simplicity, here we neglect the oxygen tilt contribution to the spontaneous strain by assuming $q_{ijkl}=0$. $\mathbf{E}_{ext}$ is the external electric field. $\mathbf{E}_d=-\nabla \phi$ is the build-in electric field evaluated from the electric potential $\phi$. The potential can be calculated from Gauss's law $\nabla\cdot\mathbf{D}=\rho(\mathbf{r})$, where $\mathbf{D}=\varepsilon_0\mathbf{E}_d+\mathbf{P}$ and $\rho(\mathbf{r})$ is the charge density. Gauss's law leads to the constraint $-\varepsilon_0\nabla^2\phi+\nabla\cdot\mathbf{P}=\rho(\mathbf{r})$. The electric potential solved in the Fourier space $\phi(\mathbf{k})$ is $\phi(\mathbf{k})=-{\rm i}\frac{k_xP_x(\mathbf{k})+k_yP_y(\mathbf{k})}{\varepsilon_0(k_x^2+k_y^2)}+\frac{\rho(\mathbf{k})}{\varepsilon_0 (k_x^2+k_y^2)}$, where $k_x$ and $k_y$ are the components of the wave vector. $P_x(\mathbf{k})$, $P_y(\mathbf{k})$ and $\rho(\mathbf{k})$ are the Fourier transformation of $P_x(\mathbf{r})$, $P_y(\mathbf{r})$, and $\rho(\mathbf{r})$, respectively.
The evolution of the AFE domain structures can be described by the time-dependent Landau Ginzburg equation
\begin{equation}
\frac{\partial\mathbf{P}}{\partial t}=-L\frac{\delta F}{\delta\mathbf{P}} \label{eq1} 
\end{equation}
where $L$ is the kinetic constant and $t$ is the time. In this work, Eq.(\ref{eq1}) is discretized using the Euler scheme on a mesh with periodic boundary conditions. We rescale the evolution time as $t^*=tL\tau_0$ ($\tau_0=2.5\times10^7$). The space discretization step of each grid $\Delta=a_0$, where $a_0$ is the lattice constant of the primary unit cell. We use the Landau parameters as: $\alpha_1=-1.34\times10^7-\lambda/a_0^2$ VmC$^{-1}$, $\alpha_{11}=4.78\times10^8$  Vm$^5$C$^{-3}$,  $\alpha_{12}=-4.98\times10^8$  Vm$^5$C$^{-3}$,  $\alpha_{111}=5.57\times10^8$  Vm$^9$C$^{-5}$, and  $\alpha_{112}=1.10\times10^9$ Vm$^9$C$^{-5}$. The elastic constants are given as $C_{11}=15.6\times10^{10}$ Nm$^{-2}$, $C_{12}=9.6\times10^{10}$ Nm$^{-2}$, $C_{44}=12.7\times10^{10}$ Nm$^{-2}$. The electrostrictive $Q_{11}=0.048$ m$^4$C$^{-2}$, $Q_{12}=-0.014$ m$^4$C$^{-2}$, and $Q_{44}=0.065$ m$^4$C$^{-2}$. In the simulation, we use $\lambda=-0.8a_0^2\tau_0$ to study the AFE structure and choose $g=0.4\lambda a_0^2$.
   
\section{Results and Discussions}

\subsection{Domain switching and hysteresis loop versus external electric field}

\begin{figure}
	\centering
	\includegraphics[width=3.2in]{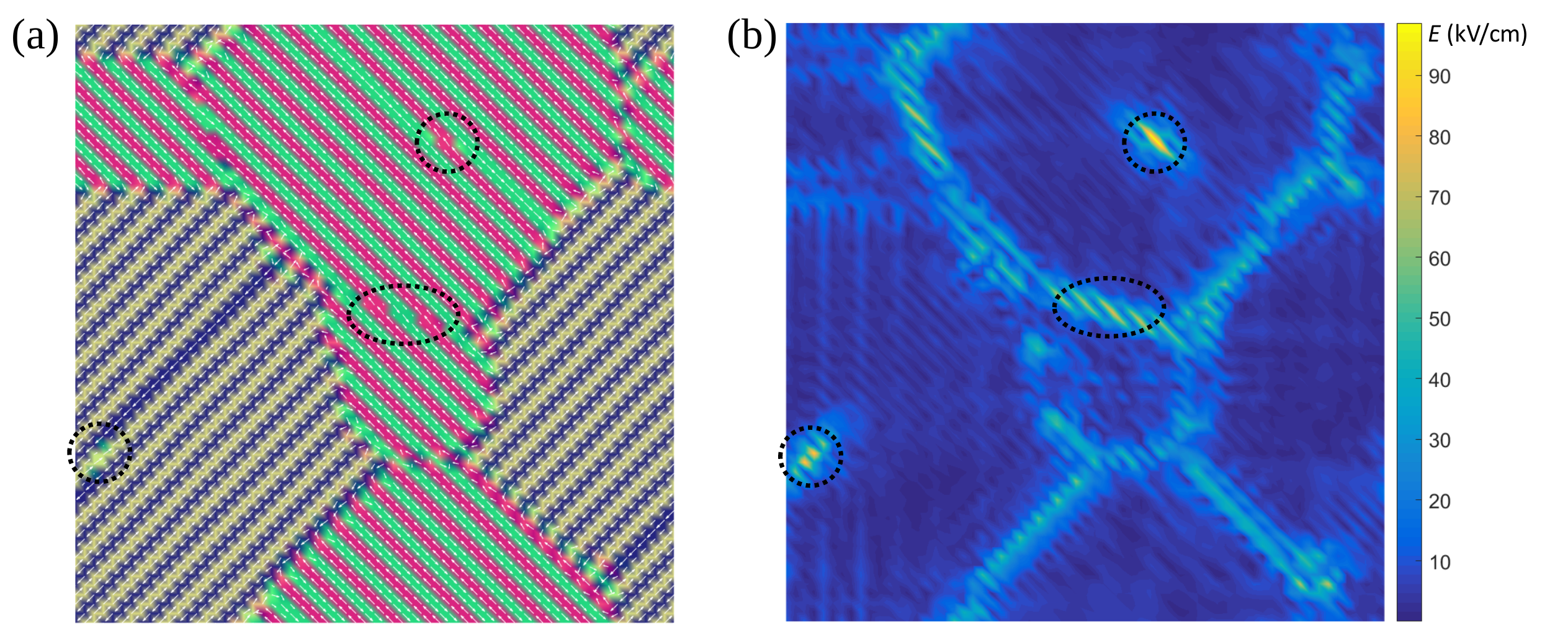}
	\caption{(a) AFE domain structure without external electric field. The white arrows represent the polarization vector, and the RGB color is used to label the orientation of the subdomains. (b) The distribution of the built-in electric field. }
	\label{fig1}
\end{figure}

\begin{figure}
	\centering
	\includegraphics[width=3.2in]{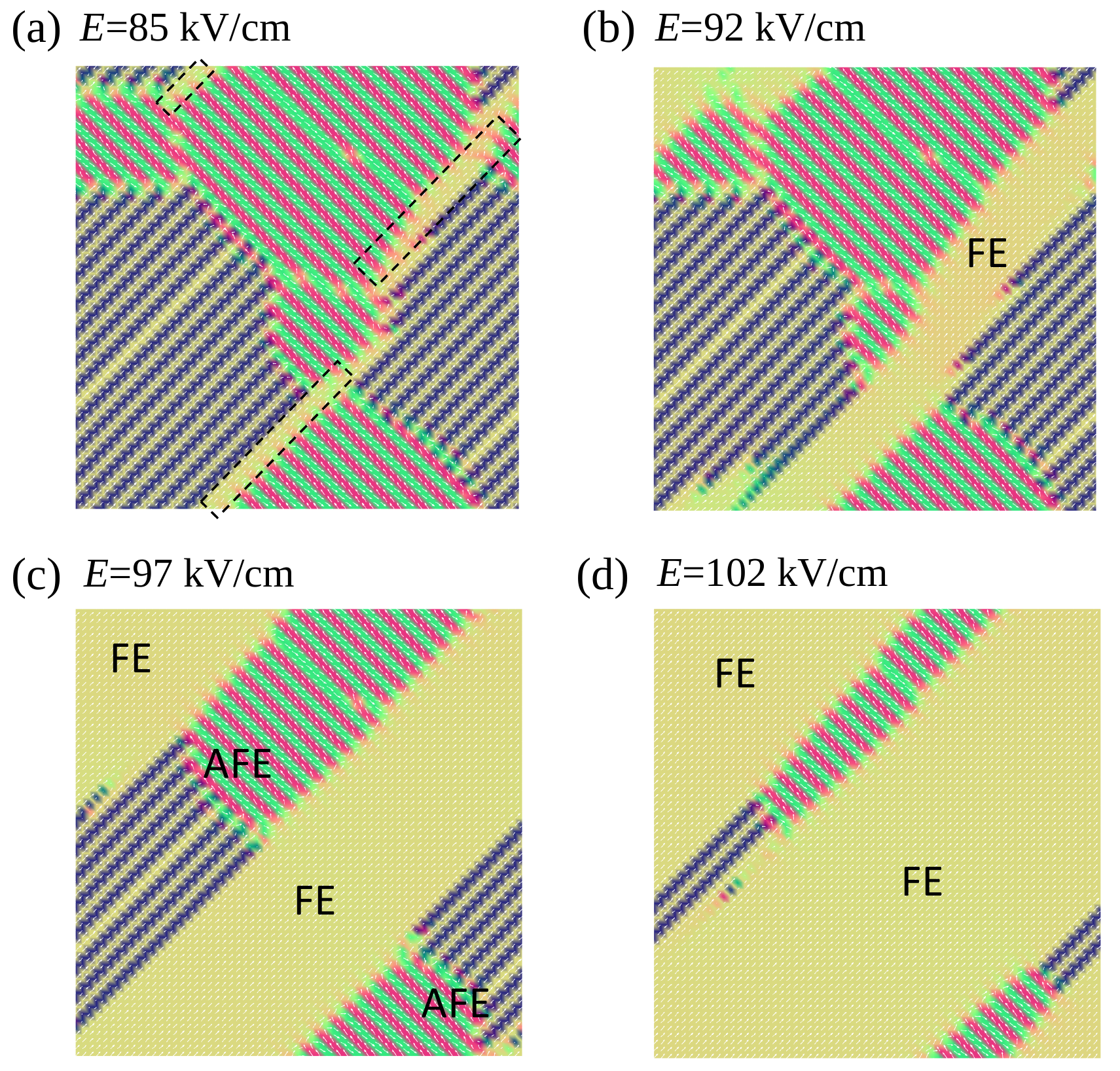}
	\caption{AFE domain switching versus external electric field along [11] direction. With the increase of applied field, the polarization start switching at the antiferroelectric domain boundaries, as labeled in the dashed line region.}
	\label{fig2}
\end{figure}

Figure \ref{fig1}(a) shows the stable domain structure which is composed of AFE domains along [11] and  [$\bar{1}1$] directions. Subdomains with anti-parallel polarization can be seen within [11] and  [$\bar{1}1$] orientated AFE domains. Although no defects are considered in the system, local dislocations could emerge in some positions at the subdomain boundaries, as shown in the dashed line region of Figure \ref{fig1}(a). Similar pattern is also confirmed in PbZrO$_3$ in experiments through STEM imaging\cite{wei2014ferroelectric}. Figure \ref{fig1}(b) shows the distribution of the built-in electric field. One can see that the magnitude of the electric field presents a higher value at the perpendicular subdomain boundaries. And the dashed line regions of Figure \ref{fig1}(b) indicate a large build-in electric field near the dislocation points.

\begin{figure}
	\centering
	\includegraphics[width=3in]{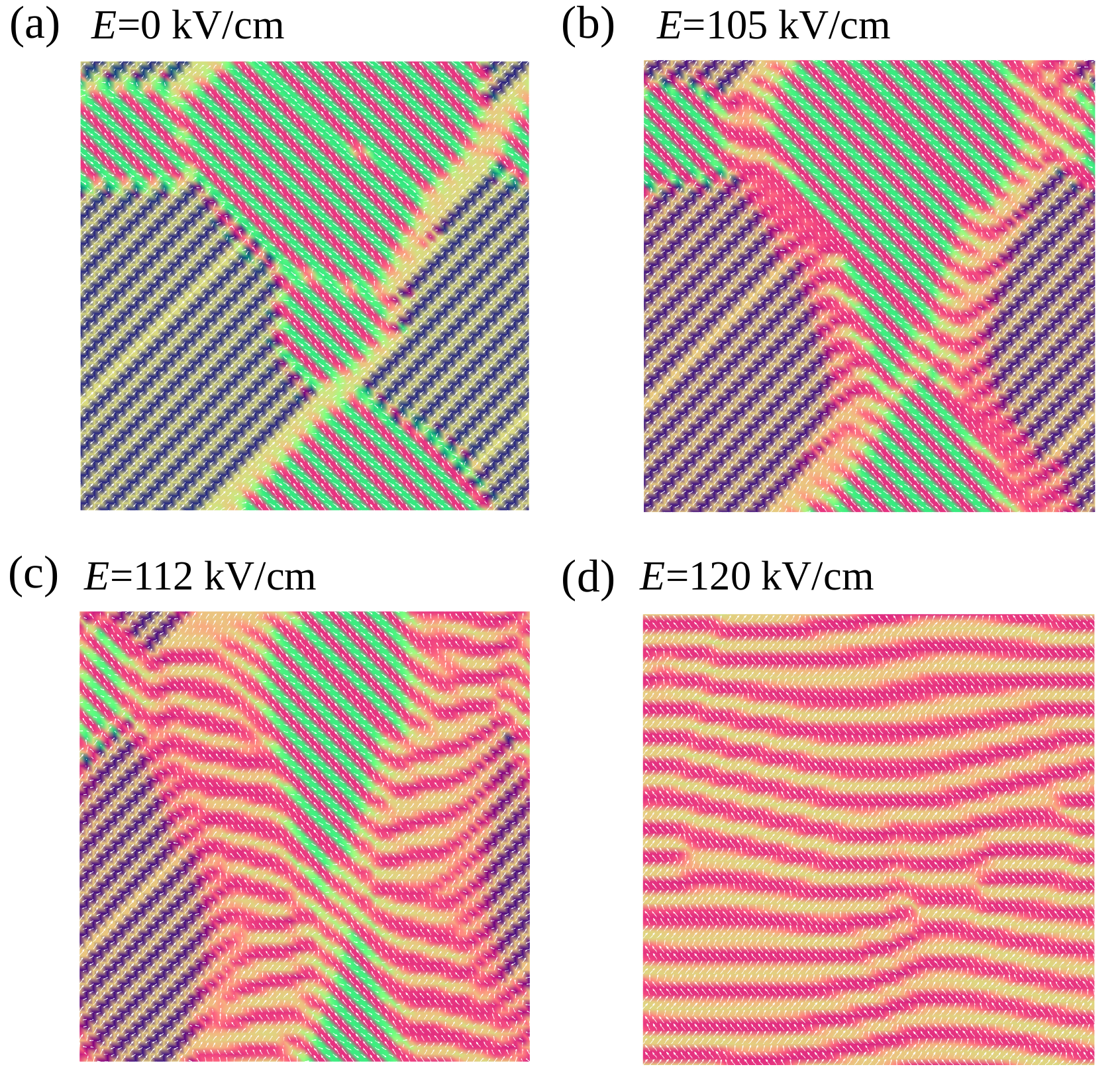}
	\caption{AFE domain switching during the increase of the external electric field along [01] direction.}
	\label{fig3}
\end{figure}

\begin{figure}
	\centering
	\includegraphics[width=3in]{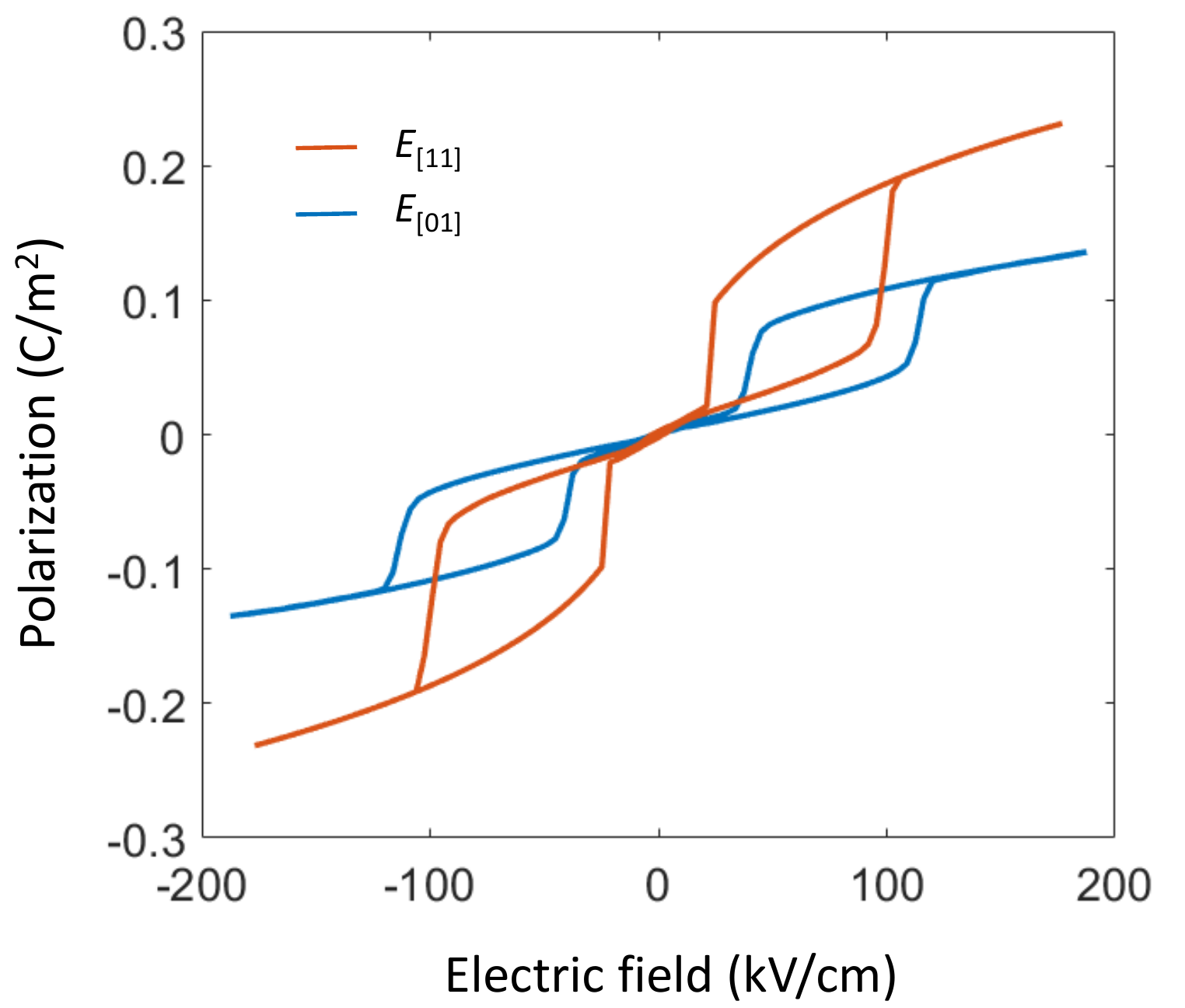}
	\caption{Calculated hysteresis loops vs the electric field along two different directions.}
	\label{fig4}
\end{figure}

\begin{figure}
	\centering
	\includegraphics[width=3in]{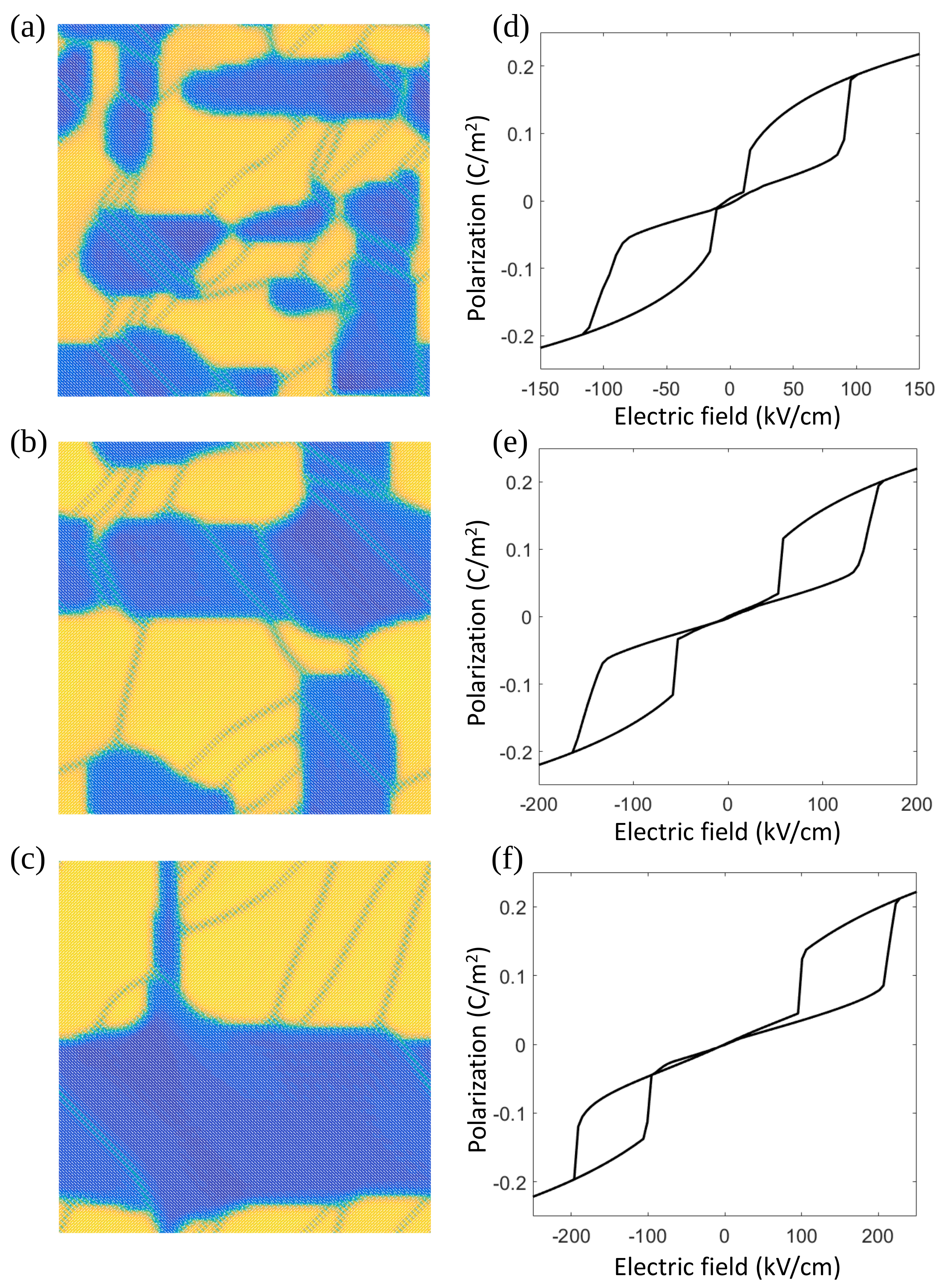}
	\caption{Size effect on the polarization switching by varying the coefficient $\lambda$. (a) $\lambda=-0.8a_0^2\tau_0$, (b) $\lambda=-1.2a_0^2\tau_0$, (c) $\lambda=-1.6a_0^2\tau_0$, (d)-(f) are the corresponding hysteresis loop of the domain structure. }
	\label{fig5}
\end{figure}

\begin{figure*}
	\centering
	\includegraphics[width=5.4in]{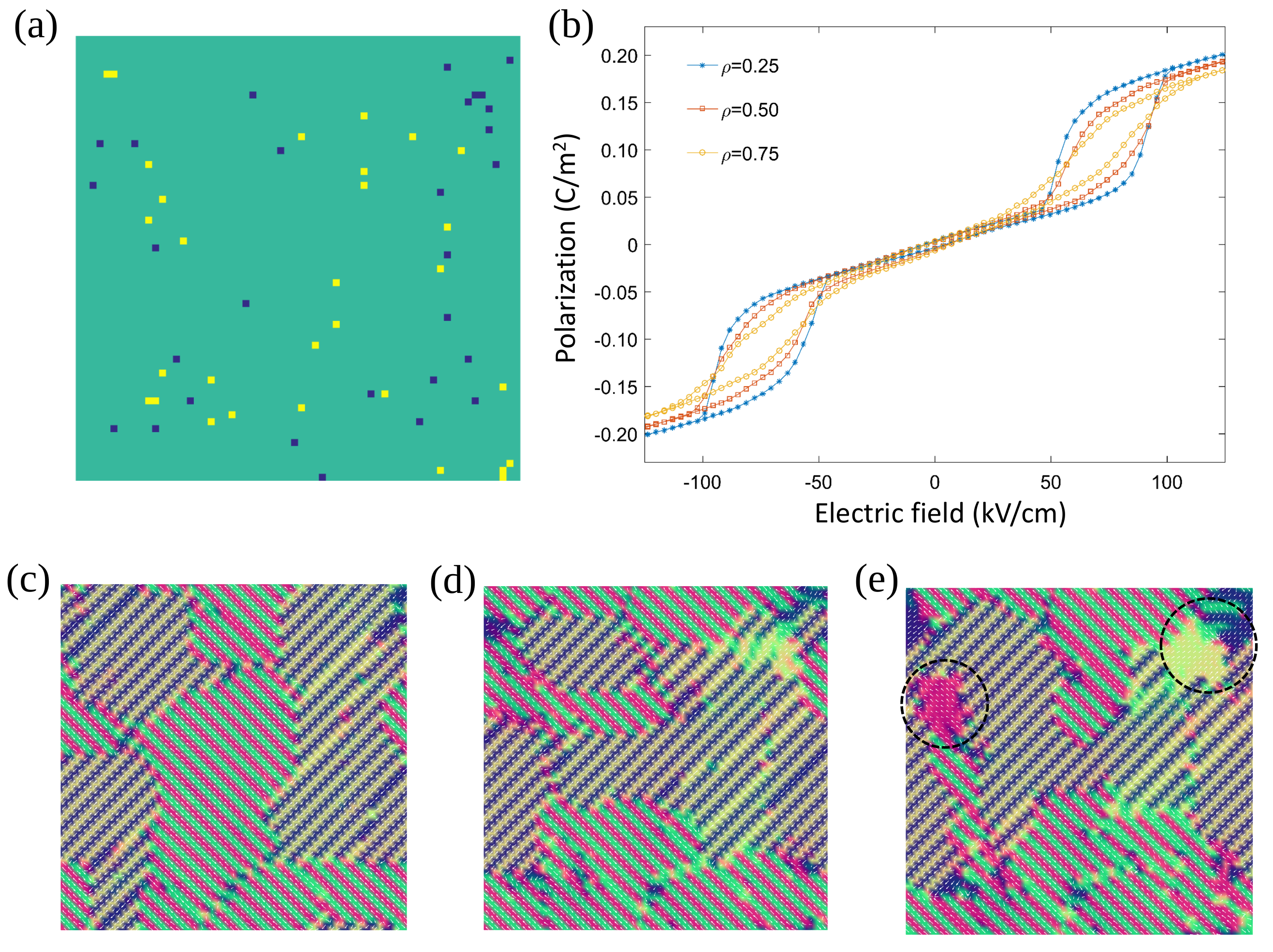}
	\caption{Influence of charge defects on AFE structures and hysteresis loops. (a) Initialized randomly distributed negative and positive charge defects in the simulation. (b) The hysteresis loops of AFE domain structures with different charge strength $\rho$. (c)-(e) Stable domain structures with different charge strengths, (c) $\rho=0.25$C/m$^2$, (d) $\rho=0.50$C/m$^2$, (e) $\rho=0.75$C/m$^2$}
	\label{fig6}
\end{figure*}

To study the polarization switching property, we have applied an external electric field along [11] and [01] direction, respectively. Figure \ref{fig2} shows the domain switching behavior under [11] electric field. At the electric field of 86kV/cm that is close to the switching  field, the polarization starts switching to the field direction at the interfaces between [11] and [$\bar{1}$1] oriented domains, and the domain structure transform from AFE to FE state, as shown in the dashed rectangular region of Figure \ref{fig2}(a). As the field keeps rising, the FE domains increase whereas the AFE domains are suppressed. In Figure \ref{fig2}(c), one can clearly see the mixture of AFE and FE phases, and the polarization arrangement on the boundary between the AFE and FE domain structures are displayed by the white arrows. Above 100kV/cm, the domain structure could be poled to a single FE domain along the field direction. Figure \ref{fig3} shows the domain switching under the electric field along [01] direction. Similarly, the polarization switching also starts at the boundaries of differently oriented domains. With the increase of the field, the AFE domains transform to FE states along [11] and [$\bar{1}$1] directions. At 100kV/cm, the AFE domains totally transform to FE phase. The average polarization versus the electric field  is calculated in Figure \ref{fig4}. It shows the saturation polarization under the [11] electric field is about 0.2C/m$^2$, which is a bit higher than 0.12C/m$^2$ of the [01] poled AFE structure. The switching electric field is about 100kV/cm in the [11] direction and 125kV/cm along the [01] direction. This result indicates that the polarization switching property is dependent on the field orientation.

For ferroelectrics, it is known that the domain size has a great influence on the properties of the materials. Nanoscale domains are more easily to switch under the external electric field, which contributes to larger piezoelectric and dielectric properties\cite{kholkin2006nanoscale,bdikin2003nanoscale,paruch2006nanoscale}. For AFEs, how the domain size influences the polarization switching is not clear yet. In the phase field simulations, we fix the Landau parameters and vary the coefficient $\lambda$ to study the domain size effect. As shown in Figure \ref{fig5}(a)-(c), with the decrease of $\lambda$ from $-0.8a_0^2\tau$ to $-1.6a_0^2\tau$, the domain size of [11] and [$\bar{1}$1] oriented (yellow and blue regions) AFE domains significantly increases. As indicated in Figure \ref{fig2}, the polarization at AFE domain boundaries is more easily to switch. Therefore, AFE structures with smaller domains can be poled to the field direction in a relatively smaller switching field. The corresponding hysteresis loops of Figure (d)-(f) indicate that larger domain size of AFE structure has no obvious influence on the magnitude of the polarization, but can greatly enhance the switching electric field.  

\subsection{Influence of charge defects}
Experimentally, it is possible that  charged defects and oxygen vacancies will stabilize in some ceramic grains or in local regions inside the crystals. For FE and AFE compositions via aliovalent doping, for instance, either a 2$^+$ or a $5^+$ ion to replace $4^+$ ion at the B-site of ABO$_3$ perovskite structure, there will be localized charge imbalance. Such charge imbalance can be compensated through A-site and oxygen vacancies, which creates many positive and negative charge defects in the material system. For the case of AFEs, charges can be also present to compensate the charged interfaces between the AFE and FE regions. The charge defects can induce a local electric field and pinning the polarization rotation. Thus it is expected to have a great impact on the polarization switching behavior of FE and AFE structures.

In order to study the influence of the charged defects on the AFE domain structure and switching property, we generate randomly distributed charge defects with a density of $\rho$ in the system, and the charge in total is electrically neutral. As shown in Figure \ref{fig6}(a), the yellow and blue points represent positive charges and negative charges, respectively. Figure \ref{fig6}(c)-(d) are the stable AFE domain structures with the increasing of charge strength $\rho$. For low charge strength with $\rho=0.25$C/m$^2$, the charge defects effect is slight that does not obviously affect the AFE domain structure. With $\rho$ increase to 0.5C/m$^2$, one can see that nanoscale FE domains start forming in the system, as shown in Figure \ref{fig6}(d). When $\rho=0.75$ C/m$^2$, many local FE domains emerge in the system due to the charged defects induced electric field. The influence of the charged defects on the hysteresis loop is calculated in Figure \ref{fig6}(b), it shows that with the increasing of charge density, the saturation polarization is slightly suppressed and the enclosed area of the hysteresis loop also decreases. The decrease of the saturation polarization is not preferred for higher energy storage density, however, the narrowed area of the hysteresis loop could give rise to  larger storage efficiency.

\section{Conclusions}

In conclusion, we have used phase field simulation to investigate the polarization switching behavior of AFE domain structures. It is demonstrated that the polarization switching to FE state starting at the boundaries of the AFE domains, and larger domain sizes could lead to a higher switching electric field. The charge defects in the materials could contribute to local FE domains in the AFE structures, and the results indicate that the appearance of charge defects in the system will suppress the saturation polarization and narrow the enclosed area of the hysteresis loop, which can enhance the energy storage efficiency while slightly reducing the storage density.

\begin{acknowledgements}
This work was supported by the LOEWE program of the State of Hesse, Germany, within the project FLAME (Fermi Level Engineering of Antiferroelectric Materials for Energy Storage and Insulation Systems).
\end{acknowledgements}

The data that support the findings of this study are available from the corresponding author upon reasonable request.

\bibliography{reference}

\end{document}